\documentclass[11pt,nofootinbib,showpacs]{revtex4}
\usepackage{amsfonts,amssymb,amsmath,graphicx}
\def\[{\left[}
\def\]{\right]}
\def\({\left(}
\def\){\right)}

\newcommand{\const}{\mathop{\rm const}\nolimits}
\begin{document}

\baselineskip7mm
\title{Note on the properties of exact solutions in Lovelock gravity}

\author{Sergey A. Pavluchenko}
\affiliation{Instituto de Ciencias F\'isicas y Matem\'aticas, Universidad Austral de Chile, Valdivia, Chile}

\author{Alexey .V. Toporensky}
\affiliation{Sternberg Astronomical Institute, Moscow State University, Moscow, 119992 Russia}

\begin{abstract}

We study the properties of cosmological solution for a flat multidimensional anisotropic Universe in Lovelock gravity.
A particular attention is paid to some features of solutions in a general Lovelock gravity which have no their
counterparts in analogous solutions of General Relativity. We consider exponential and so called generalized
Milne solutions and discuss reason for these solutions to exist in Lovelock gravity and do not exist in General Relativity.

\end{abstract}


\maketitle

\section{Introduction}

The Lovelock gravity \cite{Lovelock} being a natural generalization of General Relativity (GR) with equations of motion of the same order
as in GR became recently a matter of intense investigation mainly due to popularity of higher-dimensional paradigm.
For any fixed number of spatial dimensions the Lovelock Lagrangian includes finite number of terms (in contrast to
string gravity where we have an infinite raw of curvature corrections). The Einstein-Gilbert Lagrangian is the first
term in this theory, and this term is the only one in (3+1) dimensions. Higher order terms can be important in
multidimensional scenarios. The second term is the famous Gauss-Bonnet (GB) combination, and in (4+1) and (5+1)
dimension there are no other terms. 

The system containing  Einstein-Gilbert and Gauss-Bonnet terms have been studied in many papers during at least
last 30 years (see e.g. \cite{d4, raz1, Deruelle1, Deruelle2, raz2}). In particular, many works have been devoted to cosmological dynamics near a cosmological singularity
where quadratic in curvature contribution from Gauss-Bonnet term should be important. Moreover, it is possible to assume
that in high-curvature regime the contribution from Einstein-Gilbert term is negligible, and to consider a pure
Gauss-Bonnet gravity. Equations of motion simplify considerably under this assumption, and it is possible to find a number of
analytical solutions. 
Addtionally, if we are looking for solution in the power-law form (and we are), it is impossible to construct it with a mixture of different Lovelock contribution in the time-independent 
form (since different Lovelock contributions scale differently in time: Einstein-Hilbert are $\propto t^{-2}$, Gauss-Bonnet are $\propto t^{-4}$ and so on). 

The equations of motion as well as the resulting dynamics near a singularity for Gauss-Bonnet in (5+1) in a flat 
anisotropic Universe have been obtained previously~\cite{Deruelle1,Deruelle2, LP1} and generalized in \cite{my_prd} for a general Lovelock gravity in various dimensions. Further studies including presence of 
matter or/and a subdominant Einstein-Gilbert term have been done in \cite{Ronid, Elizalde, mpla, Kirnos1, Kirnos2,Chirkov}. 

The equations of motion for the Gauss-Bonnet gravity take the form

\begin{equation}
\sum\limits_{i>j>k>l}^D H_i H_j H_k H_l = 0~~\mbox{(constraint equation)};
\label{GB_constr}
\end{equation}

\begin{equation}
\sum\limits_{j\ne i} \( \dot H_j + H_j^2  \) \sum\limits_{\{k > l\} \ne \{ i, j\}} H_k H_l + 
3 \sum\limits_{\{a>b>c>d\}\ne i}H_a H_b H_c H_d=0 ~(i\mbox{th dynamical equation}),
\label{GB_dyn}
\end{equation}

\noindent where  $H_i \equiv \dot a_i/a_i$ is the Hubble parameter and $a_i \equiv a_i(t)$ is the scale factor 
corresponding to the $i$th coordinate.

For  the power-law {\it ansatz} ($a_i(t) = a_i^0 t^{p_i}$, $H_i(t)=\dot a_i(t)/a_i(t) = p_i/t$, $\dot H_i(t) = -p_i/t^2$) equations (\ref{GB_constr}) and (\ref{GB_dyn}) take the form:
 
\begin{equation}
\sum\limits_{i>j>k>l}^D p_i p_j p_k p_l = 0~~\mbox{(constraint equation)};
\label{GB_constr1}
\end{equation}

\begin{equation}
\sum\limits_{j\ne i} \( p_j^2 - p_j  \) \sum\limits_{\{k > l\} \ne \{ i, j\}} p_k p_l + 
3 \sum\limits_{\{a>b>c>d\}\ne i}p_a p_b p_c p_d=0 ~(i\mbox{th dynamical equation}).
\label{GB_dyn1}
\end{equation} 
 
\noindent In Gauss-Bonnet gravity one of two conditions should be applied to the power indices:

\begin{equation}
\sum\limits_{i=1}^D p_i =3 \quad\mbox{or} \quad \sum\limits_{j>k>l}^D p_j p_k p_l =0,
\label{pow_cond}
\end{equation}

\noindent where we call the first of them as the generalized Kasner solution and  the second one as the generalized Milne solution for brevity.  The power indices for the generalized Milne 
solution can be expressed in the form $(a,b,0,0,0)$ where
$a$ and $b$ are arbitrary numbers. This fact for (5+1)-dimensional case was demonstrated in \cite{mpla} and later in \cite{my_prd} was generalized for general Lovelock case.
There are no analogs of such solution in GR. On the other hand, generalized Kasner
solution looks similar to  the power-law flat anisotropic solution in GR (``classical'' Kasner 
multidimensional solution).

If only the highest order Lovelock term is considered, these results can be generalized  \cite{my_prd}.
It is interesting that though the ``classical'' Kasner solution
can be considered as a particular case in this scheme, it has some properties which separate it from any other Lovelock cases.
In the present paper we discuss several features which are specific for the Kasner solution and do not shared by its 
analogs in Lovelock gravity.

The structure of the manuscript is as follows: in the Section 2 we prove the
noncompactness of the power indices space in GB and link it with the existence of the exponential solutions. In Section 3 we discuss the properties of the generalized Milne solution. Finally in the Conclusions we outlook briefly our results.

\section{Noncompactness of the power indices space in the general Lovelock case}

Regarding generalized Kasner solution it is possible to note that despite first Kasner relation is not altered
seriously (the sum of power indices is equal to some odd number depending on the order of the Lovelock term),
the simple form of the second relation (the sum of indices squared is equal to unity) does not survive in higher order 
Lovelock theories \cite{Deruelle1, Deruelle2,LP1}. The Kasner sphere of Einstein theory is replaced by some less simple surface which appears to 
be noncompact.


First we proof the noncompactness of Kasner solutions space for the case with even number 
of spatial dimensions. The simplest case is (4+1)-dimensional Gauss-Bonnet; the Kasner
conditions are governed by two equations: $\sum_{i=1}^{4} p_i = 3$ and $p_1 p_2 p_3 p_4 = 0$.
From the second of them it is clear that one of $p_i$ should be zero while others compose a
plane that obey $\sum_{j\ne i; j=1}^{4} p_j = 3$. The complete set joins four planes according
to four different $p_i$ which can be equal to zero.

The general Lovelock case is similar to the Gauss-Bonnet one; the Kasner conditions are~\cite{my_prd}:
$\sum_{i=1}^{2n} p_i = (2n-1)$ and $p_1 p_2 \dots p_{2n} = 0$. Similarly, from the second of
them one of $p_i$ should be zero while others compose a hyper-plane that obey 
$\sum_{j\ne i; j=1}^{2n} p_j = (2n-1)$. And again, the complete set joins $2n$ hyper-planes
according to $2n$ choices of $p_i$.


The proof for the case of the odd number of spatial dimensions is a bit more complicated;
again, first we give a proof for the simplest ((5+1)-dimensional Gauss-Bonnet) case and then
extend it onto a general case.

In (5+1)-dimensional Gauss-Bonnet universe the generalized Kasner conditions are $\sum_{i=1}^{5} p_i = 3$ and
$\sum_{i>j>k>l} p_i p_j p_k p_l = 0$. 
 Let us express one of $p_i$ from the first
equation (without loss of generality let it be $p_5$), substitute it into the second equation and resolve
the result with respect to another $p_i$ (let it be $p_4$):

\begin{equation}
p_4^2 (p_1p_2 + p_1p_3 + p_2p_3) - p_4 \left( (3-p_1-p_2-p_3) (p_1p_2 + p_1p_3 + p_2p_3) 
\right) - p_1p_2p_3 (3-p_1-p_2-p_3).
\label{5dkasner}
\end{equation}

We need to show that there exist a solution of this equation when at least one of the power indices from
remaining three is arbitrary large.
Let us denote one of $\{p_1, p_2, p_3\}$ as $A$ (nonzero and arbitrary large) (let it be $p_3$); then 
the discriminant $D$ of (\ref{5dkasner}) in leading on $A$ order takes a form 
$D = A^4 (p_1 + p_2)^2 + O(A^3)$. It is obvious that if $p_1 \ne -p_2$ then 
$D > 0$ so the solution always exists (we do not consider the $p_1 = -p_2$ case since we are only interested
in noncompactness and not in some exact solutions and with $p_1 \ne -p_2$ noncompactness is already proven). 

It is worth to note that in the GR case the second bracket in the second term of the Eq. (6) is absent, and $p_3$ enters 
linearly in the second term while still quadratically in the third term, so the asymptotic of the discriminant in the
limit $p_3 \to \infty$ is no longer valid.

Now we prove our statement  for a  general case with odd number of spatial dimensions. The generalized Kasner
conditions for a general case are~\cite{my_prd}: $\sum_{i=1}^{2n+1} p_i = (2n-1)$ and 
$\sum_{i_1>i_2> \dots > i_{2n}}^{2n+1} p_{i_1} p_{i_2} \dots p_{i_{2n}} = 0$. Similarly
to (5+1) case we express $p_{2n+1}$ from first equation, substitute it into the second equation, solve
the result with respect to $p_{2n}$ (one can verify that it is still a  quadratic equation),
denote $p_{2n-1}$ as $A$ and write down the discriminant in leading order in $A$:

\begin{equation}
D = A^4 \left( \sum\limits_{i_1>i_2>\dots > i_{2n-3}}^{2n-2} p_{i_1} p_{i_2} \dots p_{i_{2n-3}} \right)^2 + O(A^3).
\label{det_2n}
\end{equation}

\noindent Similarly to (5+1) case, if the multiplier at $A^4$ is nonzero, then the solution
always exists, and for the same reason we do not consider the case when the multiplier is equal to zero.


It is interesting that power-law {\it ansatz} is not a unique possibility for a flat Universe in Lovelock gravity.
Substituting $H_i=\const; \dot H_i=0$ it is possible to get solutions corresponding to exponentially increasing
or decreasing scale factors \cite{iv1,iv2}. Such solutions are absent in GR -- formal substitution of this {\it ansatz}
only lead to trivial solution with all $H_i\equiv 0$.
One can note an interesting link: in GR we have compact power indices space ($\sum p_i^2 = 1$) and no exponential solutions. On the contrary, in GB case we have noncompact power indices space and 
there are exponential solutions. It is reasonable to think  that existence of exponential solutions is linked with noncompactness of the power indices space. Indeed, from the definition of 
$p_i = - H_i^2/\dot H_i$ (where $p_i \ne 0$)
one can see that the compactness of $p_i$ space ($|p_i| \leqslant 1$) in GR means that $|\dot H_i| \geqslant H_i^2$. This assures that $\dot H_i$ cannot be nullified  without putting to zero 
the corresponding $H_i$, making exponential solution non-existent. However, in GB case we have noncompact $p_i$ space and $\dot H_i$ can be arbitrary small in modulus. Formally, from the 
definition $p_i = - H_i^2/\dot H_i$ one can see that zeroth $\dot H_i$ (with nonzero $H_i$) corresponds to infinite $p_i$, which could be achieved only if the space of possible $p_i$ is noncompact.

\section{Properties of the generalized Milne solution}

We remind a reader that the generalized Milne solution in Gauss-Bonnet gravity is a power-law solution with indices
$(a,b,0...0)$ with only two non-zero indices which can be absolutely arbitrary. To shed a light on this rather particular solution
 we go back to the $(H_i, \dot H_i)$ coordinates (this can be done from the beginning,
however, historically this solution have been obtained in power-law {\it ansatz} which obscure its real meaning).  Hubble parameters, associated with zero
power indices are zeros, while those associated with arbitrary power indices appears to be  arbitrary functions. All equations 
of motion nullify\footnote{actually the
statement is even stronger -- ``all individual terms in all the equations of motion nullify''}, making Hubble parameters 
completely unconstrained. 
This situation have not been remarked previously, because the Milne solution have been usually obtained after imposing
the power-law {\it ansatz}, and in its power-law form the values of indices $a$ and $b$ can be fixed by initial
conditions. However, we can see easily that setting three Hubble functions to zero in enough for satisfying 
the equations of motion -- two remaining Hubble functions remain  unconstrained.
We treat this situation as an unphysical one, and we find the meaning behind it as follows: Milne solution is some kind of ``artifact'' which remains in the 
system if we neglect lower-order contribution: dealing with pure Gauss-Bonnet gravity we indeed neglect lower-order Einstein-Hilbert contribution. As on the solution under consideration all terms originating from "dominating" Gauss-Bonnet
combination vanish, neglecting Einstein-Gilbert term is obviously incorrect.

The analog of Milne solution in GR is Taub~\cite{Taub} solution, it imply $(p_1, p_2, p_3) = (1, 0, 0)$, but it is not a full analog of the Milne solution. One can easily see that 
the Taub solution is a particular case of the Kasner solution
 -- formally it follows $\sum p_i = \sum p_i^2 = 1$. A "true" GB analog of generalized Milne solution in GR
would be $(p_1, p_2, p_3) = (a, 0, 0)$, but the equations of motion require $a\equiv 1$.
The formal reason is that in the GR case the combination $(\dot H_i + H_i^2)$ is not a multiplier, so if even all other Hubble
vanish, this combination requires $p_i=1$.

If we consider not only the Gauss-Bonnet contribution, but also the Einstein-Hilbert part, and impose generalized Milne conditions, the Gauss-Bonnet contribution
will vanish and 
the constraint equation for Einstein-Hilbert contribution takes the form $H_1H_2 = 0$ 
(all other Hubble parameters are already set to zero) which imply one of $\{H_1, H_2\}$ (say, $H_2$) is also always equal to zero. After that the only term that remains in the dynamical
equations is $(\dot H_1 + H_1^2)$ which is also should be equal to zero, which leads to $p_1 = 1$ -- the Taub solution, mentioned above.

The structure of  this  reductions is similar for  higher-order corrections as well -- imposing generalized Milne of the order $n$ we nullify the $n$th order contribution and leaving $2n-2$ nonzero
power indices~\cite{my_prd}. If there is only one additional next lower order Lovelock correction, then one additional Hubble parameter is also set to zero (as it happened with the Einstein-Gauss-Bonnet case),
however, this is not enough to nullify this next order Lovelock contribution, which gives us well defined equations of motion. 
 If there are more than one additional next lower order Lovelock corrections, then no additional nullification of Hubble parameters occur.
For example, in (7+1) with cubic and quadric (Gauss-Bonnet) terms only by imposing third order generalized Milne we set three Hubble parameters to zero, which nullify cubic Lovelock contribution. Remaining 
four Hubble parameters act as effective (4+1) Gauss-Bonnet model and we have one additional Hubble parameter nullified 
(the generalized Kasner solution requires one of remaining Hubble parameters to vanish in this case, see above).
 However, if we have originally in (7+1) the linear Lovelock 
contribution as well (the Einstein-Hilbert action) then the remaining four Hubble parameters effectively form (4+1) Einstein-Gauss-Bonnet model and no additional nullification occurs. 
In both cases we obtain well-defined solutions with no arbitrary functions.




\section{Conclusions}               

In this paper we considered the dynamics of a flat anisotropic Universe in Lovelock gravity and described two situation
in which corresponding behavior in GR is different from any theory with higher order Lovelock terms.

First, we demonstrate that unlike $N$-dimensional GR where the space of possible Kasner indices represented by $(N-2)$-dimensional 
sphere, in the general Lovelock theory it is noncompact. It is interesting to link this fact with existence of exponential
 solutions in the theory under consideration and absence of those in GR.

Secondly, we discuss the nature of exceptional solution (which we call as generalized Milne solution here), which does not exist
in  GR where two Kasner conditions are the only conditions for a power-law solution.  On the contrary, in $n$th order 
Lovelock theory setting large enough number of Hubble parameters to zero results in vanishing of all terms in equations
of motion identically, leaving the rest of Hubble functions absolutely unconstrained (we even need not to impose a power-law
{\it ansatz} here!). It is reasonable to treat this situation as an artifact of neglecting lower-order Lovelock contribution.
As such solution does not exist in GR, retaining Einstein-Gilbert contribution destroys it.

\section{Acknowledgments}

This work was supported via RFBR grant No. 11-02-00643.  S.A.P. was supported by FONDECYT under Project No. 3130599.

\end{document}